\documentclass[sigconf,screen]{acmart}
\usepackage{enumitem}

\AtBeginDocument{%
  }

\newboolean{showcomments}
\setboolean{showcomments}{true}

\ifthenelse{\boolean{showcomments}}
{\newcommand{\nbc}[3]{
 {\colorbox{#3}{\bfseries\sffamily\scriptsize\textcolor{white}{#1}}}
 {\textcolor{#3}{\sf\small$\blacktriangleright$\textit{#2}$\blacktriangleleft$}}
 }
}
{\newcommand{\nbc}[3]{}
}

\copyrightyear{2025} 
\acmYear{2025} 
\setcopyright{cc}
\setcctype{by-nc-sa}
\acmConference[KDD '25]{Proceedings of the 31st ACM SIGKDD Conference on Knowledge Discovery and Data Mining V.2}{August 3--7, 2025}{Toronto, ON, Canada}
\acmBooktitle{Proceedings of the 31st ACM SIGKDD Conference on Knowledge Discovery and Data Mining V.2 (KDD '25), August 3--7, 2025, Toronto, ON, Canada}
\acmDOI{10.1145/3711896.3736572}
\acmISBN{979-8-4007-1454-2/2025/08}

\begin{document}

\title{The Hitchhikers Guide to Production-ready Trustworthy Foundation Model powered Software (FMware)}

% \affiliation{%

\author{
    Kirill Vasilevski\textsuperscript{1},
    Benjamin Rombaut\textsuperscript{1},
    Gopi Krishnan Rajbahadur\textsuperscript{1,2},
    Gustavo A. Oliva\textsuperscript{1,2},\\
    Keheliya Gallaba\textsuperscript{1},
    Filipe R. Cogo\textsuperscript{1},
    Jiahuei (Justina) Lin\textsuperscript{1},
    Dayi Lin\textsuperscript{1},
    Haoxiang Zhang\textsuperscript{1},
    Bouyan Chen\textsuperscript{1},
    Kishanthan Thangarajah\textsuperscript{1},
    Ahmed E. Hassan\textsuperscript{2},
    Zhen Ming (Jack) Jiang\textsuperscript{3}
}

\affiliation{
    \institution{\textsuperscript{1}Centre for Software Excellence, Huawei Canada; \textsuperscript{2}Queen's University, Canada; \textsuperscript{3}York University, Canada}
    \city{}
    \country{}
}

% \affiliation{
%     \institution{\textsuperscript{1}Centre for Software Excellence, Huawei Canada}
%     \city{}
%     \country{}
% }
% \affiliation{
%     \institution{\textsuperscript{2}Queen's University, Canada}
%     \city{}
%     \country{}
% }
% \affiliation{
%     \institution{\textsuperscript{3}York University, Canada}
%     \city{}
%     \country{}
% }

%\authornote{Equal contribution between Kirill V., Gopi R., Gustavo O., and Ben R.}

\renewcommand{\shortauthors}{Vasilevski et al.}

\begin{abstract}
Foundation Models (FMs) such as Large Language Models (LLMs) are reshaping the software industry by enabling FMware, systems that integrate these FMs as core components. In this KDD 2025 tutorial, we present a comprehensive exploration of FMware that combines a curated catalogue of challenges with real-world production concerns. We first discuss the state of research and practice in building FMware. We further examine the difficulties in selecting suitable models, aligning high-quality domain-specific data, engineering robust prompts, and orchestrating autonomous agents. We then address the complex journey from impressive demos to production-ready systems by outlining issues in system testing, optimization, deployment, and integration with legacy software. Drawing on our industrial experience and recent research in the area, we provide actionable insights and a technology roadmap for overcoming these challenges. Attendees will gain practical strategies to enable the creation of trustworthy FMware in the evolving technology landscape.
\end{abstract}




\maketitle
\section{Introduction}
% introduce fmware
%The increasing capabilities of Large Language Models (LLMs) led to an explosion of new type of software called FMware where foundation models (FMs) are one of the building blocks. Use of FMs opens new doors to non-experts to develop new FMware for a variety of use cases, thanks to FMs natural language interface, high flexibility, and general capabilities.

%The natural language interface of FMs allows for a large set of capabilities and high flexibility with processing input and outputs, however, it also presents several challenges when integrating with classical software.

% challenges in fmware development

%

% TODO: rephrase a bit
%We propose a lecture-style tutorial for KDD 2025, where we explore insights, concerns, and practical strategies for development and real-world productization of FMware. We first provide an overview of current state of FMware in research and practice to lay the background for the paper. We then present and discuss a number of challenges in FMware development such as model and data selection, prompt engineering, and agentic frameworks. Following that, we address the complexity of productization of FMware focusing on testing, deployment, optimization, and integration. Lastly, we will outline actionable insights for solving many of the discussed challenges.

Foundation Models (FMs), particularly Large Language Models (LLMs), are transforming software engineering by enabling \textbf{FMware}, i.e., software systems that integrate FMs as core components. FMware expands software capabilities, allowing domain experts to develop AI-driven applications without extensive programming expertise. As adoption accelerates, the market for FMware is projected to grow at a compound annual rate of 35.9\% through 2030~\cite{llmmarket}. Despite this rapid growth, developing FMware remains a significant challenge, requiring engineering practices that differ fundamentally from traditional software development.

Building FMware is difficult because it introduces stochastic behavior, evolving dependencies, and high computational costs while lacking established engineering methodologies. Developers must handle unpredictable outputs, data sensitivity, prompt brittleness, and limited interpretability, making even reliable prototyping difficult. FMware often integrates multiple AI components, creating interdependencies that amplify failure points and complicate engineering. These complexities set FMware apart from conventional software, demanding new software engineering approaches.

Deploying FMware at scale presents an even greater challenge. Production-ready FMware must ensure performance stability, robustness, and compliance while managing scalability constraints, security risks, and deployment costs. Industry experiences illustrate these difficulties: LinkedIn required four additional months to refine the final 20\% of an FMware deployment, facing diminishing returns on each incremental improvement~\cite{linkedinMusingsBuilding}. Microsoft and GitHub reported that as FMware complexity increases, validation and testing become prohibitively expensive~\cite{parnin2023building} due to the absence of standardized evaluation frameworks. The cost of deployment further complicates production readiness. For instance, OpenAI’s ChatGPT infrastructure alone was estimated to cost \$700,000 per day in 2023~\cite{uniteFinancialChallenges}.

Existing research has explored FMs' role in software engineering, including their applications in code generation, testing, documentation, and reasoning~\cite{jin2024llms, fan2023large, zhang2023survey, hou2024large, jiang2024survey}. These studies primarily rely on academic benchmarks and structured datasets, often overlooking real-world deployment challenges. Meanwhile, research on software production readiness has largely focused on release criteria, system reliability, and ML-specific deployment challenges~\cite{asthana2009quantifying, cusick2013architecture, breck2017ml, parnin2023building, oreillyWhatLearned}. Studies on enterprise FM integration~\cite{nahar2024beyond, chen2025empirical} highlight developer concerns but remain constrained to specific ecosystems. A recent survey accompanying a KDD 2024 tutorial~\cite{ding2024reasoning} examined how LLMs enhance software development workflows. While that work focused on improving development processes using LLMs, we examine a different challenge:\textbf{ the engineering complexities of FMware itself}. Unlike prior surveys, which emphasize how LLMs assist software engineering, we analyze the core software engineering difficulties of building, deploying, and maintaining FMware at scale. Furthermore, this survey improves on previous work \cite{rajbahadur2025cooldemosproductionreadyfmware, hassan2024rethinking} by providing a cohesive overview of all challenges related to FMware development in a way that is relevant to researchers and practitioners alike. Our work systematically identifies these challenges and provides a roadmap for addressing them.
% \gopi{Might have to adjust wording later}

To bridge these gaps, we propose a KDD 2025 lecture-style tutorial that explores insights, concerns, and practical strategies for developing and deploying trustworthy, production-ready FMware. 
% This survey paper accompanies the tutorial, providing a structured analysis of the field.
We first present an overview of the FMware lifecycle in Section~\ref{sec:lifecycle}. We then discuss key the challenges and present a critical overview of the state of practice in FMware engineering in Section~\ref{sec:challenges}. Finally, we outline actionable insights and emerging solutions, offering a roadmap for advancing FMware engineering in Section~\ref{sec:roadmap}. Section~\ref{sec:conclusion} concludes our paper.

% \kirill{removed last paragraph to save space}
% By synthesizing recent research, industry case studies, and hands-on development insights, we provide a comprehensive perspective on the engineering challenges of FMware and the pathways to trustworthy, production-ready solutions. As FMware adoption accelerates, the need for AI-native software engineering methodologies becomes increasingly urgent. We aim to equip researchers and practitioners with the strategies needed to bridge the gap between experimental prototypes and scalable, real-world AI systems.\gopi{Not sure if this paragraph is needed or not, but leaving it there for now}

\section{FMware Lifecycle}~\label{sec:lifecycle}
\begin{figure*}
  \centering
  \includegraphics[width=1.0\textwidth]{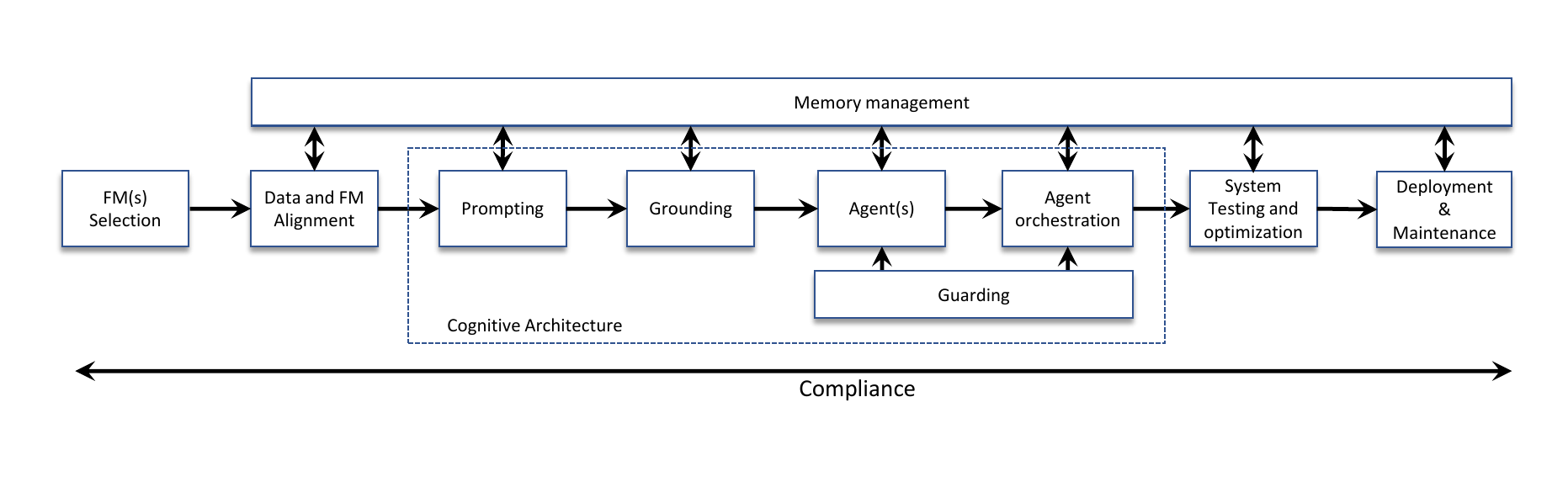}
  \caption{FMware Lifecycle.}
  \label{fig:lifecycle}
\end{figure*}

The FMware lifecycle defines how foundation models (FMs) are integrated into software systems, from selection and alignment to long-term operation. Figure~\ref{fig:lifecycle} shows an overview of the FMware lifecycle. Unlike traditional software, which follows deterministic execution paths, FMware relies on probabilistic inference, dynamic adaptation, and real-time decision-making. This requires new engineering approaches that blend AI-driven capabilities with structured software development principles. Below we discuss about the different stages of FMware lifecycle. %and the state-of-practice in the industry and research. 
% We are also provide a critique at times where the

\subsection{FM Selection and Alignment}
\label{sec1:model_alignment}

Selecting an appropriate FM is the first step in the FMware lifecycle. Unlike traditional software components with predefined logic, FMs generate responses based on statistical patterns from large-scale training data. Developers must consider task requirements, scalability, licensing constraints, and adaptability when choosing an FM.

Once selected, alignment is necessary to fine-tune an FM for application-specific needs. \textit{Data alignment} involves curating domain-relevant datasets, ensuring the FM performs reliably on specialized tasks~\cite{liu2024best, albalak2024survey}. \textit{Model alignment} further refines behavior using techniques such as instruction tuning and reinforcement learning with human feedback (RLHF) to guide model outputs toward desired behaviors~\cite{shen2023large, xia2024understanding}.
\begin{comment}
\noindent \textbf{State of Practice.}  
To reduce manual effort in alignment, researchers explore \textit{active learning}~\cite{yu2022actune,toloka} and \textit{weak supervision}~\cite{biegel2021active, bach2019snorkel}. Approaches like Snorkel AI~\cite{ratner2017snorkel} and Self-Instruct~\cite{wang2023selfinstruct} generate labeled datasets automatically, though manual oversight is still required~\cite{maekawa2022low}. Recent studies focus on \textit{data quality over quantity}, showing that fine-tuning on high-quality instructional data reduces alignment costs~\cite{gunasekar2023textbooks, zhou2023lima}. 

To improve debugging, researchers employ \textit{influence functions} and \textit{feature importance analysis} to identify impactful training data points~\cite{grosse2023studying, schoch2023data}. While traditional data testing tools~\cite{braiek2020testingml, krishnan2016activeclean} aid dataset cleaning, they struggle with the high dimensionality and opacity of FMs, making fine-grained debugging an ongoing challenge.
\end{comment}

\subsection{Prompt Engineering and Grounding}

Since FMware interacts with users through natural language prompts rather than explicit function calls, prompt engineering defines system behavior. Developers iteratively refine prompts to improve accuracy, consistency, and adaptability across different contexts~\cite{parthasarathy2024ultimate}. Production-ready systems require structured prompt design, incorporating formatting, context augmentation, and constraint mechanisms.

To enhance reliability, developers employ well-established \textit{prompt patterns}, such as chain-of-thought prompting~\cite{chainOfThought2022}, self-consistency prompting, and structured templating. These patterns improve coherence, reduce variability, and enable more predictable FM responses. Prompt portability is another critical consideration, ensuring that prompts remain effective when transitioning between different FMs~\cite{hassan2024rethinking}. However, manually crafting and optimizing prompts remains a labor-intensive and iterative process, leading researchers and practitioners to explore automated and systematic solutions.

\textbf{Grounding} improves response reliability by linking FM outputs to external information sources. Techniques such as retrieval-augmented generation (RAG) ensure factual accuracy and reduce hallucination risks~\cite{lewis2020rag}. Grounding also aids compliance by ensuring alignment with authoritative sources, mitigating risks of generating misleading content~\cite{nahar2024beyond}.
\begin{comment}
\noindent \textbf{State of Practice.}  
To address the inefficiencies in prompt development, various tools and methods have emerged. Visual tools like Prompt IDE~\cite{xPromptIDE} and Aleph Alpha~\cite{deiseroth2023atman} provide inference transparency and token-level explainability to assist FMware developers in refining prompts. However, they lack higher-level debugging capabilities for analyzing logical inconsistencies. Researchers explore semi-automated approaches~\cite{hsieh2023automatic, xu2022gps, ye2023prompt} and reinforcement learning-based optimization~\cite{deng2022rlprompt} to improve prompt effectiveness. For example, Automatic Prompt Engineer~\cite{zhou2023large} and PromptBreeder~\cite{fernando2023promptbreeder} use evolutionary search techniques to generate optimal prompts, while DSPy~\cite{khattab2023dspy} integrates prompt optimization with a programming model. However, these techniques often optimize prompts in isolation, ignoring variations in model candidates, inference parameters, and hardware environments, which are critical in real-world FMware deployment.

\end{comment}

\subsection{Agents, Orchestration, and Cognitive Architecture Planning}

While FMware can operate through direct prompt-response interactions, modern applications increasingly incorporate autonomous agents for decision-making, task execution, and interoperability with external systems. Agents extend FM capabilities by enabling multi-step reasoning, maintaining state, and integrating with traditional software components. An FMware agent functions as an adaptive entity that executes complex workflows, interacts with external tools and databases, and refines decision-making through feedback loops.

To manage these capabilities, cognitive architecture planning defines how FMware components coordinate within a system and enable agent orchestration. Depending on the application, orchestration can take different forms: \textbf{static workflows}, where execution paths are predefined; \textbf{tool-augmented execution}, where FMs interact with databases or predefined functions dynamically; and \textbf{agent-based orchestration}, where autonomous agents adjust execution flows and collaborate with other agents in real time.

FMware agents integrate with traditional software architectures, enhancing existing deterministic systems by bridging structured programming with adaptive reasoning.

\begin{comment}
\noindent \textbf{State of Practice.}  
To facilitate multi-generational software integration, current practices encapsulate legacy systems as tools that FMware can access. Frameworks like Semantic Kernel~\cite{semantickernel} use plugins and \emph{native functions} to abstract legacy system capabilities into modular components. AutoGen~\cite{wu2023autogen} and Copilot Workspace~\cite{githubnextcopilotworkspace} offer mechanisms for human oversight, allowing users to guide agent behavior. MetaGPT~\cite{hong2023metagpt} encodes standard operating procedures (SOPs) into prompts, while symbolic planning approaches~\cite{zhou2023agents} enhance controllability in agent-driven systems. These methods highlight the importance of balancing automation with human oversight to ensure reliable FMware orchestration.
\end{comment}

\subsection{Observability and Guarding}

Observability in software engineering refers to the ability to monitor system behavior, detect failures, and analyze performance trends. Traditional observability relies on logs, traces, and metrics to ensure expected functionality. However, FMware introduces new challenges due to its probabilistic nature, requiring deeper insights into dynamic behaviors beyond conventional monitoring.

Semantic observability extends traditional practices by capturing user corrections, intent recognition, and model drift detection to refine FMware performance over time. Unlike classical observability, which tracks system health at a functional level, FMware observability must also analyze emergent behaviors arising from model interactions and decision-making processes.

Ensuring trust in FMware requires robust guarding mechanisms to enforce compliance, safety, and reliability~\cite{rebedea2023nemo, dong2024framework}. Unlike static rule-based validation, FMware must assess correctness dynamically at inference time, incorporating safeguards for data validation, response filtering, and risk mitigation.
\begin{comment}
\noindent \textbf{State of Practice.}  
Efforts to enhance FMware observability are advancing but remain largely adapted from traditional software monitoring. Tools like OpenLLMetry~\cite{traceloopOpensourceObservability} and LangSmith~\cite{langchainLangSmith} capture traces of FM calls, vector database interactions, and latency metrics. However, they lack insights into the cognitive processes underlying FMware decisions, especially in multi-agent environments where behaviors emerge dynamically. GitHub Copilot employs semantic signal telemetry by tracking user edits after code generation, but generalizable solutions for FMware-wide semantic monitoring are still lacking.

More sophisticated frameworks, such as Microsoft PowerAutomate, offer workflow-level tracing, allowing developers to diagnose agent coordination issues. However, these approaches focus on functional observability rather than capturing how agents plan, reason, and interact. While LangSmith provides structured logging for LangChain-based applications, it remains limited to single-agent workflows, leaving multi-agent observability an open challenge.
\end{comment}
\subsection{Performance Engineering, Testing, and Optimization}

FMware validation differs from traditional software testing due to its non-deterministic outputs. Unlike conventional applications, where the same input produces the same output, FMware responses vary based on model updates, prompt changes, and shifting input distributions~\cite{ma2024my}. Ensuring reliability requires continuous evaluation pipelines that benchmark consistency across evolving model versions while assessing distributional stability, response variability, and edge case handling~\cite{microsoftRetryStorm}. Optimization efforts extend beyond accuracy, balancing latency, compute efficiency, and cost constraints. Unlike deterministic APIs, FM inference introduces variable processing times, requiring adaptive strategies such as query batching, caching, and model routing.
\subsection{Deployment, Maintenance, and Tooling}

Deploying FMware extends beyond traditional CI/CD practices, as model behavior evolves dynamically rather than following static updates~\cite{llmpilot}. Maintaining FMware requires continuous inference monitoring, prompt tuning based on observed performance, and version rollback mechanisms to mitigate performance regressions. Unlike conventional software, FMware lacks standardized tooling for versioning, fine-tuning management, and scalable inference deployment, necessitating adaptive update strategies to ensure stability.
\begin{comment}
\noindent \textbf{State of Practice.}  
Current FMware development relies on Git for prompt and prompt template versioning, while simple agent workflows are often defined in YAML, such as in Microsoft PromptFlow~\citep{mspromptflow}. However, no established standards exist for describing prompts and agents. Initiatives like the AI Engineer Foundation’s \textit{agent protocol}~\citep{agentprotocol} and ongoing IEEE SA efforts aim to standardize agent interactions, with frameworks like AutoGPT~\citep{autoGPT} adopting similar protocols. Unlike traditional software (GitHub) and machine learning (Hugging Face), FMware lacks a central hub for sharing and collaborating on assets, limiting ecosystem-wide innovation.

Feedback integration in FMware remains rudimentary, relying on explicit user ratings~\cite{wang2021towards} rather than passive, scalable mechanisms for continuous model refinement. Production environments demand automated feedback collection to detect biases, inaccuracies, and performance shifts in real time. Existing approaches also fail to distinguish between \textit{outer knowledge} (global insights applicable across users) and \textit{inner knowledge} (user-specific optimizations), leading to either overlooked improvements or misapplied generalizations. Addressing these gaps is essential for enabling adaptive, production-ready FMware systems.
\end{comment}
\subsection{Cross-cutting concerns}
\subsubsection{Memory Management}

FMware requires efficient memory management to preserve context across interactions, optimize multi-agent workflows, and manage inference workloads~\cite{microsoftRetryStorm, luo2024arena}. Unlike traditional applications with explicitly stored state, FMware must dynamically handle context persistence, agent coordination, and compute optimization to ensure efficiency.
\begin{comment}
\noindent \textbf{State of Practice.}  
Effective memory management remains a key challenge in FMware, particularly in synchronizing knowledge across multiple FMs and user contexts. Systems often struggle with consistency in multi-feedback loops~\cite{microsoftCheckYour}, making feedback integration unreliable. During deployment, memory handling supports model swapping and prevents performance degradation from latency issues~\cite{llmpilot}. Managing retries efficiently avoids costly retry storms while ensuring agents share state and context for coherent decision-making~\cite{microsoftRetryStorm, luo2024arena}. Robust memory systems are also critical for regulatory compliance and optimizing dynamic workloads~\cite{hassan2024rethinking, wang2023survey}.

Knowledge representation inefficiencies inflate computational costs and response times by storing redundant or irrelevant data. Poor retrieval optimization can lead to inconsistent reasoning and incorrect outputs~\cite{guo2024knowledgenavigator, wu2024easily}. Additionally, managing memory across FMware instances is error-prone due to synchronization issues, conflicting knowledge, and data corruption~\cite{xie2023adaptive, chen2022rich}. Production-ready FMware requires robust memory architectures to mitigate context loss and memory mismanagement, ensuring stable and reliable performance in deployment environments~\cite{packer2023memgpt}.
\end{comment}

\subsubsection{Compliance \& Regulations}

Compliance in FMware spans the entire lifecycle, requiring continuous evaluation due to the probabilistic nature of model outputs and evolving regulations. Organizations must ensure that training data and model alignment comply with licensing, privacy laws, and ethical AI principles~\cite{hassan2024rethinking, nahar2024beyond}. Regulatory risks arise from data misuse, while model fine-tuning techniques such as RLHF must be audited to prevent biases.

Inference-time compliance demands real-time safeguards, including content filtering, bias detection, and automated moderation to mitigate harmful outputs~\cite{dong2024framework, rebedea2023nemo}. Post-deployment, FMware requires robust logging, auditing, and transparency measures such as model cards and AI Bill of Materials (AI BOM)~\cite{hassan2024rethinking}. Governance frameworks must define accountability, risk management, and alignment with emerging AI regulations~\cite{nahar2024beyond}, ensuring responsible FMware deployment while maintaining adaptability to regulatory shifts.
\begin{comment}
\noindent \textbf{State of Practice.}  
Regulatory compliance in FMware remains largely manual and process-heavy. Efforts such as enhanced model and data cards~\cite{brajovic2023model} and AI risk management frameworks~\cite{nist2021guidance, novelli2023taking} provide guidance but lack scalability for complex FMware systems. License compliance has seen advances with Responsible AI Licenses (RAIL)~\cite{contractor2022behavioral} and OpenDataology~\cite{opendataology}, but these solutions often focus on individual models or datasets rather than the integrated FMware ecosystem. Emerging standards like SPDX 3.0 Dataset Profile~\cite{SPDXDatasetProfile2023} and AI BOMs offer promising directions for automating compliance and tracking dependencies, though gaps remain in addressing compliance for synthetic data and evolving FMware components, particularly in scenarios involving agent-driven learning and adaptation.
\end{comment}

The FMware lifecycle defines how foundation models integrate into modern software systems, introducing new considerations for prompt engineering, agent orchestration, observability, and continuous adaptation. Unlike traditional software, FMware evolves dynamically through interaction feedback, semantic observability, and model-driven optimization, requiring engineering approaches that blend AI, software reliability, and real-time adaptability. As FMware adoption accelerates, structured software engineering principles will be critical to ensuring scalability, trustworthiness, and enterprise-level integration.

\section{Development \& Productionization Challenges in FMware}
\label{sec:challenges}

In this section, we critically analyze the state of practice in FMware development and productionization. We examine existing approaches, their limitations, and the persistent challenges that hinder the transition from experimental FMware to robust, trustworthy production-ready systems. 

\subsection{FM Selection and Alignment}

Current approaches to FM selection and alignment remain largely ad hoc. Selecting and aligning FMs for FMware requires balancing functionality, cost, latency, and infrastructure constraints. As of February 2025, over 1.4 million open-source models are available on Hugging Face~\cite{huggingfaceModels}, alongside proprietary models from providers like OpenAI~\cite{openaiModels} and Anthropic~\cite{anthropicModels}. Proof-of-concept FMware often relies on large general-purpose FMs (e.g., GPT-4o~\cite{openaiGPT4o}, o1~\cite{openaio1}, Claude 3.7 Sonnet~\cite{sonnet37}) due to their broad capabilities, but transitioning to production necessitates optimizing for task-specific accuracy while mitigating cost and deployment overheads~\cite{kamath2024llms, Zhao25}. Developers rely on manual experimentation to compare models, with few automated benchmarking tools for real-world deployment scenarios. Additionally, complex FMware systems may require multiple specialized models rather than a single FM, compounding the selection challenge.

Beyond model selection, data and model alignment are critical to ensuring that an FM meets specific enterprise and use-case requirements~\cite{xia2024understanding}. However, ensuring data quality, bias mitigation, and efficient data estimation, required for efficient data alignment remains a persistent challenge. Low-quality datasets degrade reliability~\cite{albalak2024survey}, while poor domain coverage can introduce unintended biases. More recently, prioritizing data quality over volume has been shown to improve alignment efficiency~\cite{gunasekar2023textbooks, zhou2023lima}. Furthermore, estimating the required data volume is difficult—underestimation leads to degraded performance, whereas overestimation results in unnecessary computational costs~\cite{xia2024understanding}.

Finally, debugging alignment remains challenging, as FM opacity limits fine-grained dataset impact analysis, even with techniques like influence functions and feature importance analysis~\cite{grosse2023studying, schoch2023data}. Automating model evaluation and selection for production FMware remains an open problem.

\subsection{Prompt Engineering and Grounding}
\label{sec:prompt_eng}

\subsubsection{\textbf{Prompt Engineering}}
Designing prompts that reliably elicit the desired behavior from foundation models (FMs) remains a significant challenge in FMware development. While basic applications rely on ad-hoc prompt crafting, production-grade FMware demands a structured approach to ensure robustness, maintainability, and cross-model portability~\cite{parthasarathy2024ultimate}. Despite efforts to automate prompt generation through optimization techniques~\cite{zhou2023largelanguagemodelshumanlevel, khattab2024dspy}, most real-world systems still rely on trial-and-error refinement, making the process time-consuming and error-prone~\cite{strobelt2023visualprompt}.

Another key challenge is \textit{prompt sensitivity}, where minor changes in wording, model updates, or inference parameters can lead to drastic output variations~\cite{gao2021makingllmsbetterfewshot}. This instability complicates software maintenance and necessitates continuous prompt adjustments. Additionally, prompts optimized for one FM often fail to generalize across different models due to architectural differences and variations in tokenization~\cite{chen2023chatgpts, sabbatella2024prompt}. Such lack of portability hinders FMware evolution and increases development costs.

Another major issue is the \textit{lack of built-in validation mechanisms}, making it difficult to assess prompt reliability. Unlike traditional software, where unit tests verify deterministic behavior, FMware lacks standardized QA methodologies to systematically test prompt performance~\cite{ahmed2023better, kamoi2024evaluating}. Debugging also remains challenging, as failures can stem from poorly designed prompts, FM limitations, or inference-time parameters~\cite{amujo2024good, ashtari2023discovery}. Research into \textit{automated prompt optimization}~\cite{fernando2023promptbreederselfreferentialselfimprovementprompt} and \textit{reinforcement learning-based prompt tuning}~\cite{deng2022rlprompt} has shown promise, but such techniques remain limited by their inability to adapt across different FMs and deployment environments.
\subsubsection{\textbf{Grounding}}
Integrating grounding into FMware presents multiple challenges, including data selection, retrieval efficiency, and content integration. A primary challenge is ensuring \textit{high information density} in grounding data. Poorly structured or overly broad information can degrade FM performance, leading to irrelevant or incoherent responses~\cite{laban2024summary}. While structured retrieval can improve precision, it requires carefully curated datasets and filtering mechanisms to maintain relevance~\cite{cuconasu2024power, liu2024lost}. Additionally, balancing retrieval breadth and response conciseness is difficult, as excessive grounding information can lead to response dilution, while insufficient data may fail to correct FM errors~\cite{wang2024adapting}.

Another challenge is optimizing \textit{retrieval efficiency}. Many FMware systems rely on computationally expensive neural retrieval methods, such as dense vector search, to improve grounding accuracy. However, these techniques often introduce latency overhead, which can hinder real-time applications~\cite{sawarkar2024blended}. In some cases, lightweight keyword-based retrieval achieves comparable performance with significantly lower computational costs, yet determining when to use neural or lexical retrieval remains an open problem~\cite{oreillyWhatLearned}.

Finally, the seamless \textit{integration of retrieved information} into FM responses is non-trivial. While naive concatenation of retrieved documents can overwhelm the model with extraneous data, overly restrictive filtering risks omitting crucial context. Effective fusion strategies, such as selective in-context learning or hybrid retrieval-conditioning approaches, are needed to improve response accuracy without introducing excessive complexity~\cite{wang2024adapting}. %Despite ongoing advancements, grounding remains a challenging aspect of FMware, requiring continued improvements in retrieval efficiency, content selection, and integration techniques to ensure reliable and verifiable model outputs.

\begin{comment}
% moved to optimization
\subsubsection{Guardrails}

Despite the growing adoption of guardrails, current implementations lack adaptability and fail to handle complex linguistic manipulation. Many rely on static keyword filtering, which is easily bypassed via paraphrasing or subtle rewording~\cite{dong2024framework}. While semantic filtering offers improvements, it struggles with contextual ambiguity, often flagging safe content or missing nuanced violations.

FMware also lacks reliable confidence estimation, making it difficult to flag uncertain outputs for review~\cite{ayyamperumal2024current}. Without built-in mechanisms to assess response reliability, human oversight remains the primary fallback, limiting scalability in large deployments.
\end{comment}

\subsection{Agents, Orchestration, and Cognitive Architecture Planning}

Agents are integral to FMware, enabling decision-making, automation, and interaction with external tools and software. However, their effective deployment in production FMware remains an open challenge, requiring careful consideration of modularity, orchestration, and tool integration.

\textbf{Agent modularity} is crucial for reliability and maintainability. While large monolithic agents provide broad functionality, they introduce compounding errors, increase debugging complexity, and reduce adaptability~\cite{chen2024coder}. Conversely, breaking down tasks into smaller, specialized agents enhances modularity and scalability~\cite{rasal2024navigating}, but also raises coordination challenges. As agent ecosystems grow, managing interdependencies, communication, and context sharing becomes increasingly difficult, requiring robust state management mechanisms.

\textbf{Agent orchestration} is another key challenge. While FM-based planners can automatically decompose tasks and assign workloads, they often suffer from hallucinations and inconsistent task delegation~\cite{guo2024large, huang2024understanding}. The inherent unpredictability of FM-driven orchestration makes it difficult to ensure reliable task execution, particularly in multi-agent FMware. Hybrid approaches that incorporate structured task delegation mechanisms, rather than relying solely on FM-based planning, are needed to improve consistency. Existing frameworks attempt to address this by introducing structured oversight. For instance, AutoGen~\cite{wu2023autogen} and Copilot Workspace~\cite{githubnextcopilotworkspace} enable human intervention in agent workflows, providing corrective oversight to mitigate FM-based errors. Similarly, MetaGPT~\cite{hong2023metagpt} encodes standard operating procedures (SOPs) into prompts, guiding agents towards structured decision-making, while symbolic planning approaches~\cite{zhou2023agents} enhance controllability.

\textbf{Tool integration} presents additional challenges. Agents rely on external tools and APIs, but their ability to correctly select and use tools depends heavily on documentation structure and abstraction levels. High-level abstractions mitigate complexity, as exposing low-level tool functions increases error rates and impairs scalability. Poorly documented or capability-centric tool descriptions further hinder FMware reliability by making it harder for agents to generalize tool usage~\cite{wang2023voyager}. Some frameworks attempt to abstract tool interactions for FMware systems. For instance, Semantic Kernel~\cite{semantickernel} encapsulates legacy software functionalities using \textit{native functions}, making them accessible to FM-driven agents while maintaining modularity.

Despite these efforts, current orchestration frameworks remain limited in handling real-time reactivity, dynamic coordination, and ensuring robustness across complex agent-driven workflows. Production-ready FMware must address these gaps by refining agent orchestration strategies, improving state-sharing mechanisms, and incorporating structured oversight to mitigate FM unpredictability.

\subsection{Performance Engineering, Testing, \& Optimization}

\subsubsection{\textbf{Performance Optimization}} 
Production FMware systems demand rigorous optimization due to the compute-intensive nature of FMs and the need to ensure Service Level Agreements (SLAs)~\cite{bommasani2021opportunities}. A primary challenge is high inference latency, which is exacerbated by long prompts, multi-round interactions, and large model sizes~\cite{jiang2023mistral}. Using external FM providers introduces additional constraints such as token limits, request throttling, and unpredictable performance variations, all of which impact enterprise FMware deployment.

One approach to improving performance is selecting smaller, more efficient models. For example, Meta prioritized the 1.3B InCoder model over a 6.7B variant to reduce inference latency while maintaining acceptable accuracy~\cite{murali2023codecompose}. Additionally, techniques such as quantization, distillation, and pruning help optimize FM inference speed, though balancing efficiency without performance degradation remains an open problem~\cite{zhu2023survey}. Semantic caching~\cite{gim2023prompt} is another promising optimization strategy, reducing redundant computation by storing and retrieving previous responses for similar queries.

Enterprise FMware often integrates multiple FMs and agents, requiring multiple rounds of inference. However, due to the non-deterministic nature of FM-based orchestration and planning, the exact number of inference rounds is often unknown before execution. This unpredictability can cause FMware to miss SLAs, even if individual FMs meet model-level constraints. This issue is magnified in agent-based FMware, where agents continuously generate new tasks and queries, making runtime costs difficult to estimate.

To address infrastructure constraints, organizations may opt to self-host FMs rather than relying on external providers. While this mitigates vendor limitations, it introduces challenges in managing hardware resources efficiently. In FMware requiring multiple models, GPU memory contention forces frequent model swaps, adding tens of seconds to inference time per operation. These swap-induced latencies necessitate optimized scheduling strategies to prevent bottlenecks in production deployments.

\subsubsection{\textbf{Testing}} 
Testing FMware is fundamentally different from traditional software testing due to its non-deterministic outputs and evolving model behaviors~\cite{parnin2023building, parry2022surveying}. Even with unchanged code, FMware responses can vary due to model updates, inference temperature changes, and external FM provider modifications. These challenges are particularly severe for systems relying on closed-source FM APIs, where developers have limited control over model internals.

Current FMware testing strategies remain immature and heavily rely on human evaluation or automated reference-based validation using large FMs such as GPT-4~\cite{liu2023gpteval}. However, FM-based evaluators exhibit only moderate correlation with human assessments (0.51)~\cite{liu2023gpteval}, often favoring superficial linguistic features over factual correctness~\cite{thakur2024judging, chen2024humans}. This inconsistency raises concerns about the reliability of AI-as-judge frameworks in evaluating FMware outputs.

Additionally, traditional software testing approaches do not fully address FMware’s dynamic nature and resource constraints. Existing caching mechanisms, such as those used in LangChain, fail to account for real-time model updates and changing prompts, leading to ineffective cache invalidation. Furthermore, FMware lacks dependency-aware test execution strategies~\cite{sharif2021deeporder}, resulting in redundant and computationally expensive testing pipelines. Retry mechanisms also remain underdeveloped; while conventional systems rely on re-execution for fault recovery, FMware often requires adaptive re-prompting or alternative query structures to ensure robustness. 

Due to the complexity of testing multi-FM and multi-agent FMware, system-wide test suites must account for dynamic interactions, robustness under workload variations, and handling of unpredictable responses. However, current automated testing frameworks remain insufficient for large-scale FMware, making fault detection slow, increasing regression risks, and limiting CI/CD capabilities~\cite{scottlogicTestingLLMBased,hassan2024rethinking}. Addressing these challenges requires new FMware-native testing methodologies, including structured prompt evaluation, dependency-aware regression testing, and improved reliability metrics for evaluating FMware correctness across inference variations.

\subsection{Observability, Guarding, Deployment, Maintenance \& Tools}
%\subsection{Observability, Guarding, Deployment \& Maintenance}

\smallskip \noindent \textbf{Limited FMware-Native Observability:} Observability in FMware remains a major challenge, as existing tools lack support for tracing reasoning paths, emergent behaviors, and inter-agent coordination~\cite{hassan2024rethinking}. While traditional observability platforms such as LangSmith~\cite{langsmith_url}, TraceLoop~\cite{traceloop_url}, and Weights \& Biases~\cite{wandb_weave_url} provide operational metrics like latency, token usage, and grounding accuracy, they fail to capture cognitive processes and implicit decision-making in FM-driven applications.

This issue is particularly problematic in multi-agent FMware, where decision-making is distributed across multiple autonomous components. Current tools do not provide insight into how agents collaborate, share state, or dynamically adjust workflows. This lack of transparency complicates debugging, increases the risk of reasoning failures, and hampers the ability to ensure system-wide reliability. Furthermore, regulatory compliance mandates detailed logging of FM interactions, yet FMware lacks standardized mechanisms for maintaining audit trails across evolving deployments.

\smallskip \noindent \textbf{Lack of Efficient Feedback Mechanisms:} Continuous feedback integration is essential for improving FMware performance, yet current implementations remain ad hoc and manual. Unlike structured feedback loops in traditional ML systems, FMware relies primarily on explicit user ratings~\cite{wang2021towards} rather than passive, scalable feedback mechanisms. While GitHub Copilot employs semantic signal telemetry (e.g., tracking post-editing behavior)~\cite{copilot_internals_url}, no generalizable solution exists for capturing implicit corrections and real-world performance issues across diverse FMware applications.

Another challenge is distinguishing global insights (outer knowledge) from user-specific optimizations (inner knowledge). Without this differentiation, valuable model refinements may be overlooked, or user-specific preferences may be misapplied across all users, leading to degraded system-wide reliability. Production FMware must incorporate \textit{automated feedback pipelines} that facilitate real-time bias detection, error correction, and adaptive learning.

\smallskip \noindent \textbf{Ineffective FM Update Mechanisms:} Unlike traditional software, where updates introduce incremental and predictable improvements, FMware updates are often non-deterministic and unpredictable~\cite{hassan2024rethinking}. Adding more training data does not guarantee resolution of specific issues, and enhancements listed in release notes often lack clear mappings to observable behavior changes. As a result, developers struggle to test, validate, and adapt to new model versions, leading to performance regressions and degraded user trust.

Additionally, FM updates introduce hidden risks causing previously stable features to break, and model alignment may shift in unintended ways. This unpredictability complicates maintenance in production environments, requiring automated monitoring and rollback mechanisms to detect and mitigate regressions effectively.

\subsubsection{\textbf{Guarding}}

Despite widespread adoption of guardrails, current implementations lack adaptability and struggle with complex linguistic manipulation. Many rely on static keyword filtering, which can be easily bypassed through paraphrasing or subtle rewording~\cite{dong2024framework}. While \textit{semantic filtering} provides more contextual awareness, it remains prone to false positives and false negatives, failing to distinguish between benign variations and genuine violations.

FMware also lacks confidence estimation mechanisms, making it difficult to flag uncertain outputs for human review~\cite{ayyamperumal2024current}. Without \textit{real-time reliability assessments}, FMware deployments must default to human oversight, which severely limits scalability in large-scale applications. Addressing these limitations requires adaptive guardrails that dynamically refine themselves based on evolving threats and real-world usage patterns.

\subsection{Cross-cutting Concerns}

% \kirill{can be cut?}
Cross-cutting concerns permeate across all stages of the FMware lifecycle, with solutions to them being essential for development of reliable production-ready FMware.

\subsubsection{\textbf{Memory Management}}

Effective memory management in FMware is essential for maintaining context consistency, optimizing workloads, and ensuring scalable inference~\cite{kwon2023efficient, hassan2024rethinking}. Unlike traditional applications with explicit state management, FMware dynamically stores and retrieves knowledge to preserve context across interactions, manage agent state, and handle retries efficiently~\cite{microsoftRetryStorm, luo2024arena}.

\smallskip \noindent \textbf{Knowledge representation inefficiencies} remain a core challenge. Poorly structured memory leads to redundant or irrelevant data storage, inflating computational costs and degrading response quality~\cite{guo2024knowledgenavigator, wu2024easily}. Optimizing knowledge retrieval is essential to \textit{prevent} inconsistencies in multi-step reasoning and minimize unnecessary computation overhead.

\smallskip \noindent \textbf{Knowledge synchronization} across multiple FMs and user contexts is another persistent issue. Feedback loops can introduce conflicting knowledge, leading to misalignment between FM states, context loss, and incorrect outputs~\cite{xie2023adaptive, chen2022rich}. Managing retry storms, state sharing across agents, and ensuring consistent knowledge updates is particularly challenging in FMware where decisions are probabilistic and evolve dynamically~\cite{microsoftCheckYour, llmpilot}.

\smallskip \noindent Additionally, robust memory architectures are vital for regulatory compliance and optimizing long-term FM behavior~\cite{hassan2024rethinking, wang2023survey}. Without structured memory management, FMware remains prone to context fragmentation, inefficient adaptation, and unpredictable system performance in production deployments~\cite{packer2023memgpt}.

\subsubsection{\textbf{Compliance \& Regulations}}

Regulatory compliance in FMware is complex due to data privacy concerns, opaque licensing, and lack of transparency mechanisms. Unlike traditional software, FMware’s probabilistic nature requires ongoing compliance assessment, as model behaviors evolve dynamically rather than remaining static.

\smallskip \noindent \textbf{Data privacy regulations}, such as GDPR, require enterprises to prevent sensitive data leakage, posing challenges when FMware relies on third-party models hosted externally~\cite{parnin2023building}. Self-hosting FMs mitigates this risk but is often infeasible due to infrastructure and resource constraints.

\smallskip \noindent \textbf{License compliance} remains a critical challenge, particularly in closed-source FMs where data provenance is undisclosed, raising legal and ethical concerns regarding the use of copyrighted or restricted datasets~\cite{shi2023detecting,longpre2023data,chang2023speak,mozes2023use}. Existing AI licensing frameworks, such as Responsible AI Licenses (RAIL)~\cite{contractor2022behavioral} and SPDX 3.0 Dataset Profiles~\cite{SPDXDatasetProfile2023}, attempt to address these gaps but remain limited to individual models or datasets, failing to account for FMware-wide compliance across multiple integrated components.

\smallskip \noindent \textbf{Lack of transparency and oversight} is another regulatory barrier, particularly under frameworks like the EU AI Act, which mandates clear explainability and human oversight for high-risk AI systems~\cite{edwards2021eu}. However, FMware’s black-box nature makes achieving these requirements intractable with current methodologies. While efforts like enhanced model and data cards~\cite{brajovic2023model} and AI Bill of Materials (AI BOMs)~\cite{opendataology} aim to improve traceability, they lack standardization and do not fully capture the evolving nature of FMware components.

\smallskip \noindent Despite progress in compliance tooling, current approaches remain manual, fragmented, and poorly scalable for enterprise FMware. Without automated compliance monitoring and adaptive governance frameworks, regulatory risks will continue to challenge FMware’s adoption in high-stakes domains.

\section{Action Plan}
\label{sec:roadmap}

Driven by the surveyed challenges listed in Section~\ref{sec:challenges} as well as our practical experience in shipping real-world FMware, we henceforth present an \textit{action plan}. This action plan is organized into a broad set of directions, each supplemented by a concise list of directives. While some directives can be somewhat readily applied to mitigate challenges, others are research topics in themselves. Therefore, we see this action plan as a valuable instrument to bridge research and practice (a key goal of KDD).

\subsection{FM Selection and Alignment}

\noindent \textbf{Systematic choice of models.}  Investigate new or improved benchmarking suites that go beyond single-metric leaderboards (e.g., include cost-efficiency parameters). Create analytical and simulation-based frameworks to explore how different models impact cost, latency, and accuracy.

\noindent \textbf{Data IDEs Covering the Entire Alignment Lifecycle.} Integrate model- and human-in-the-loop approaches for labeling, debugging, testing, and versioning data. Incorporate programmatic labeling (data programming) to reduce manual effort. Enable collaborative review and aggregation of labeled data from multiple labelers.

\noindent \textbf{Bias Detection, Compliance, and Testing Frameworks.} Develop methods to automatically identify and mitigate inconsistencies or biases in alignment data. Provide mechanisms for ensuring regulatory and ethical compliance across different stages of data handling. Facilitate testing and validation of aligned data to improve quality and trustworthiness.

\noindent \textbf{Asset Management for Open and Inner Source Data}. Offer robust management and tracking capabilities to address licensing restrictions and maintain compliance. Ensure data accuracy, consistency, completeness, and representativeness for diverse tasks. Mitigate risks associated with shared or publicly available data through secure handling and version control.
    
\noindent \textbf{User-Centric Data Programming and Labeling.} Simplify labeling workflows and define user-friendly labeling functions to increase adoption and reduce errors. Support collaboration between technical and non-technical stakeholders (e.g., domain experts) for more effective data curation. 

\subsection{Prompt Engineering and Grounding}

\noindent \textbf{Multi-Prompt Architectures and Human-in-the-Loop Workflows.} Develop prompt chains with branching (e.g., chain-of-thought, tree-of-thought) to handle complex tasks. Incorporate human feedback at intermediate steps to refine the final output. Establish design, debugging, visualization, and evolution techniques for increasingly complex cognitive processes.
    
\noindent \textbf{Integrated Prompt IDEs for Collaboration and Quality Assurance} Support real-time collaboration, including version control and role-based access for prompt design. Provide just-in-time quality assurance with prompt debuggers, linters, and visualization of prompt performance. Enable dynamic selection and adjustment of lower-level prompts based on model responses, context, and user interactions.
    
%\noindent \textbf{Reproducibility and Plug-and-Play Architectures} Explore mechanisms to ensure full reproducibility beyond simply setting the model temperature to zero. Offer a plug-and-play ecosystem that integrates community-contributed tools and services for prompt creation and execution. Facilitate near real-time monitoring and adjustment of prompt optimization strategies.

\noindent \textbf{Asset Management and Knowledge Integration.} Provide collaborative environments for sharing, reviewing, and refining prompts at global or project-specific levels. Incorporate debugging features that reveal token-level or component-level contributions to generated outputs. Integrate grounding data management (e.g., uploading, tokenizing, vectorizing, and retrieving data) to ensure high-quality and contextually accurate responses.

\subsection{Agents, Orchestration, and Cognitive Architecture Planning}

%\noindent \textbf{Multi-Generational Integration for FMware} Develop mechanisms for integrating with legacy systems to address both evolution and compatibility concerns. Embrace \textit{reusability} and \textit{modularity} by encapsulating shared functionalities into separate modules (e.g., skills) and unifying them under well-defined interfaces (e.g., agents). Simplify cross-generational maintenance by isolating legacy-specific behaviors and ensuring clear boundaries for integration.

\noindent \textbf{Fine-Grained Controllability and Component-Based Design} Adopt established software engineering principles (e.g., \textit{single responsibility} and \textit{modularity}) to decompose FMware into small, self-contained components with fine-grained control boundaries. Introduce permission models and constraints to regulate component interactions, preserving user privacy and security. Provide holistic architectural views that help identify design deviations and maintain cohesion, coupling, and separation of concerns in FMware.

\noindent \textbf{Built-In Quality Through Structured Knowledge and Curriculum} Move beyond simple vector databases toward knowledge graphs for richer, more structured contextual information. Implement \textit{curriculum engineering} \cite{wang2024curriculum, hassan2024ainativesoftwareengineeringse} and versioning to ensure agents build compositional skills and reuse existing knowledge effectively. Automate curriculum co-creation and continuous QA to remove redundant or outdated information, thereby improving accuracy, reliability, and maintainability of FMware. Employ semantic and structural checks for prompt-like inputs to reduce errors, hallucinations, and unpredictable outputs.

\noindent \textbf{Controlled Execution and Systematic Validation} Design frameworks that enforce \textit{repeatability} by restoring execution snapshots and monosemantic units to reproduce exact conditions for debugging and regression testing. Allow \textit{guided exploration} of alternative execution paths, varying inputs and agent decisions in a controlled manner to uncover hidden bugs and assess system robustness. Balance consistent execution flows with exploration to optimize both reliability and resilience of FMware applications.

\subsection{Observability and Guarding}

\noindent \textbf{Multi-Level Observability Framework for FMware.} Capture both functional performance metrics and internal cognitive processes across various abstraction layers. Provide traceability at the system level and throughout autonomous agent decision-making stages.

\noindent \textbf{Analytics Engine and Visualization Tools.} Implement a “flight recorder” mechanism to selectively log agent reasoning and communications. Integrate analytics engines for real-time visualization of multi-agent workflows, quickly pinpointing root causes in complex scenarios. Mitigate the observer effect (e.g., altering behavior by prompting “think step-by-step”) through surrogate agents that enable debugging without interfering directly.
   
\noindent \textbf{Advanced Telemetry and Metrics.} Develop enhanced, context-aware metrics to help developers assess capabilities, analyze behaviors, and validate performance in dynamic FMware environments. Employ semantic signal telemetry solutions that can generalize beyond specific implementations or domains. Offer monitoring, logging, and alerting systems tailored to the unique operational requirements of FMware.

\subsection{Compliance}

\noindent \textbf{Automated FMwareBOM Generation and Formal Verification.} Extend SPDX 3.0 (AI and dataset profiles) to create a comprehensive FMwareBOM for components, licenses, synthetic data, RLHF data, and user feedback. Employ formal verification (e.g., SMT solvers) to ensure provable compliance with legal and regulatory requirements. Integrate FMwareBOM tools into the development lifecycle to facilitate real-time, machine-readable compliance checks.
    
\noindent \textbf{FMware-Specific Licensing and Compliance Tooling} Develop tools similar to Fossology, tailored to detect and manage licensing issues unique to FMware (e.g., code, dataset, FM, and agent licenses). Establish documentation standards and disclosure procedures for compliance verification, including known limitations. Enable automated reasoning on generated FMwareBOMs for preliminary license and regulatory compliance assessments.

%\noindent \textbf{Privacy and Data Protection Mechanisms.} Implement safeguards at every level to protect enterprise and personal data, enforcing adherence to relevant privacy regulations. Balance the need for comprehensive telemetry data with the minimization of sensitive information exposure. Employ privacy-preserving techniques (e.g., data obfuscation, partial encryption) to maintain visibility into FMware behavior without disclosing confidential information.

\subsection{Performance Engineering, Testing, and Optimization}

\noindent \textbf{Declarative Representation for Intent-Aware Optimization.} Represent FMware at a higher level of abstraction, preserving developer intent (e.g., data locality and task placement considerations). Enable application graph-level optimization by capturing the execution graph in a declarative format. Facilitate joint optimizations between front-end (intent-aware representation), back-end (graph performance), and FM inference accelerations.

\noindent \textbf{Resource-Aware QA Framework and Smart Caching.} Integrate a robust FMware-native caching mechanism to store and reuse FM query results, reducing repeated FM calls and maintaining cache integrity. Dynamically prioritize and optimize test execution, grouping similar tests to lower resource consumption and accelerate feedback. Leverage prompt compression techniques to further reduce latency and resource demands during testing.
    
\noindent \textbf{Dynamic Prompt Re-Engineering and Retry Mechanisms.} Implement advanced retry strategies that adapt prompts and validate responses, reducing the risk of repeated failures. Adjust queries intelligently when outputs are unsatisfactory to improve the likelihood of success in subsequent attempts. Balance thorough testing with resource efficiency to keep FMware systems robust and performant.

\noindent \textbf{Automated Test Generation Through Metamorphic Relations (MRs).}  Leverage real-world user feedback (e.g., thumbs-up/down) to generate domain-specific metamorphic relations. Conduct metamorphic testing (MT) on FMware to evaluate diverse system properties and behaviors. Integrate an iterative feedback loop where experts refine the MRs, enhancing test coverage and reliability.
    
\noindent \textbf{Next-Generation AI-Judge Framework.} Create specialized AI-judges capable of evaluating deeper attributes such as factual accuracy, consistency, and compliance with domain logic~\cite{lin2024engineeringaijudgesystems}. Enable developers to craft prompts for these judges that capture specific business constraints, moving beyond superficial checks. Employ curriculum engineering techniques to train judges in a structured, guided, and flexible manner for cost-effectiveness and adaptability.
    
\noindent \textbf{Reduced Human Effort with Enhanced Reliability.} Streamline test result analysis and summarization through automation, reducing dependence on heavy manual evaluation. Combine automated testing with AI-judges to address the non-deterministic nature of FM outputs. Evolve testing mechanisms with each iteration, continuously enhancing the robustness of FMware quality assurance.

\subsection{Deployment, Maintenance, and Tooling}

\noindent \textbf{Efficient Deployment.} Consider FM characteristics, hardware heterogeneity, and data movement costs for global optimizations in deployment, rather than treating FM inference as a standalone concern. Simplify computing resource management through \textit{uni-clusters}, unifying training, fine-tuning, serving, and agent customization in a single infrastructure.

\noindent \textbf{Automated Feedback Integration.} Passively collect implicit user signals (e.g., hesitations, corrections) to capture behavioral cues, not just explicit ratings. Employ real-time analysis of user interactions to continuously refine and improve FMware responses. Reduce reliance on manual, explicit feedback by seamlessly integrating AI-driven feedback solicitation into regular usage patterns.

\noindent \textbf{Knowledge Partitioning and Federated Learning.} Distinguish between “outer knowledge” (general improvements) and “inner knowledge” (user-specific optimizations) to prevent overgeneralization. Automate the categorization of feedback into these knowledge domains for more targeted adaptation. Incorporate federated learning techniques to share “inner knowledge” across multiple FMware instances, enhancing overall performance without compromising privacy.

\noindent \textbf{Versioning and Collaboration Support for FMware Assets} Define standards and protocols to version-control fine-grained FMware entities (e.g., individual prompt components like persona, instructions, examples). Facilitate inter-organizational reuse through community-driven platforms and prompt hubs, promoting asset sharing and interoperability. Leverage role-based access control to maintain security and granular management of asset visibility and contributions. 

%\noindent \textbf{Unified Platform and End-to-End Lifecycle Integration.} Develop a single, flexible environment that supports the entire FMware lifecycle (from design and testing to deployment and continuous improvement). Offer seamless extensibility so teams can mix-and-match capabilities while maintaining consistent governance and process. Enable cradle-to-grave orchestration for FMware projects, reducing siloed tooling and ensuring a cohesive development experience.

% \subsection{Memory Management and System State}

% \gus{I don't know what to write here.} \kirill{imo, can remove, there's enough insights}
\section{Conclusion}~\label{sec:conclusion}

%The field of FMware offers an immense potential with organizations slowly starting to tackle the challenges of developing complex and production-ready FMware. This survey and accompanying tutorial for KDD 2025 presents an overview of the current state of FMware landscape followed by real-world development and production challenges that FMware developers will face. Furthermore, we provide a detailed action plan that aims to help FMware developers overcome these challenges. 
As FMware adoption grows, bridging the gap between research and production remains a challenge due to scalability, reliability, and a plethora of other issues that we detail in this paper. Our tutorial along with this survey paper provides a structured perspective on FMware’s lifecycle, integrating insights from AI, data science, and software engineering to address practical deployment challenges. By highlighting key gaps in alignment, orchestration, observability, and optimization, it equips the KDD community with strategies for building scalable, trustworthy, and production-ready FMware.

% old
% \input{introduction.tex}
% \input{background.tex}
% \input{req-engineering.tex}
% \input{code-implementation.tex}
% \input{testing.tex}
% \input{maintainance.tex}
% \input{document.tex}
% \input{other_research.tex}
% \input{conclusion.tex}

% \begin{acks}

% \end{acks}

\bibliographystyle{ACM-Reference-Format}
\bibliography{main}

% \appendix


\end{document}